\documentclass[runningheads,a4paper]{llncs}

\usepackage{amssymb}
\usepackage{graphicx}

\usepackage{url}
\newcommand{\keywords}[1]{\par\addvspace\baselineskip
\noindent\keywordname\enspace\ignorespaces#1}

\begin{document}

\mainmatter  

\title{Cascade Failures from Distributed Generation in Power Grids}

\titlerunning{Cascade Failures from Distributed Generation in Power Grids}

%
%
\author{Antonio Scala\inst{1,2,3}
\and Sakshi Pahwa\inst{4} \and \\
Caterina Scoglio\inst{4} }
\authorrunning{Antonio Scala et al.}

\institute{
ISC-CNR Physics Dept., Univ. "La Sapienza" Piazzale Moro 5, 00185 Roma, Italy
\and IMT Alti Studi Lucca, piazza S. Ponziano 6, 55100 Lucca, Italy
\and London Institute of Mathematical Sciences, 22 South Audley St
\\Mayfair London W1K 2NY, UK
\and Department of Electrical and Computer Engineering,  
College of Engineering  
Kansas State University, Manhattan, KS
}

%
%

\maketitle

\begin{abstract}
Power grids are nowadays experiencing a transformation due to the introduction of Distributed Generation based on Renewable Sources. At difference with classical Distributed Generation, where local power sources mitigate anomalous user consumption peaks, Renewable Sources introduce in the grid intrinsically erratic power inputs. By introducing a simple schematic (but realistic) model for power grids with stochastic distributed generation, we study the effects of erratic sources on the robustness of several IEEE power grid test networks with up to $2 \times 10^3$ buses. We find that increasing the penetration of erratic sources causes the grid to fail with a sharp transition. 
We compare such results with the case of failures caused by the natural increasing power demand.
\end{abstract}

\section{Introduction}

The unavailability of massive and economic power storage does not allow yet 
to integrate the stochastic and often volatile renewable sources 
by leveraging on distributed storage \cite{Baghaie2010}. The power-on-demand 
paradigm upon which power grids have been originally engineered did not 
contemplate the introduction of Distributed Generation from renewable sources;
at the same time, power grids are nowadays required to be robust and smart, 
i.e. to be able to maintain, under normal or perturbed conditions, the frequency 
and  amplitude supply voltage variations into a defined range and to provide fast 
restoration after faults.

In order to understand how to ensure stability and avoid loss
of synchronization, most studies have concentrated on the dynamic behavior
of Smart Grids during typical events like the interconnection of
distributed generation. To study such events in grids with a large number of elements,
simplifications like the mapping among the classic swing equations \cite{StaggBOOK1968} 
and Kuramoto models \cite{Filatrella2008,Fioriti2009,DorflerSIAM2012} 
are welcome in simplifying the numerical and the analytical study of the synchronization 
and the transient stability of a power network.

Beside dynamic instabilities, new kind of failures are possible as 
Smart grids are going to insist on pre-existing networks 
designed for different purposes and tailored on different paradigms and : 
simple models \cite{DobsonHICSS2001} akin to the DC power flow model \cite{WoodBook1984} 
show that the network topology can dynamically induce complex blackout size probability 
distributions (power-law distributed), both when the system is operated near its 
limits \cite{CarrerasCHAOS2002} or when the system is subject to erratic 
disturbances \cite{SachtjenPRE2000}. In general, new realistic metrics to assess 
the robustness of the electric power grid with respect to complex events 
like cascading failures \cite{YoussefITCP2011} are needed.

A further effect of the introduction of Distributed Generation from renewable sources 
is the possibility of an erratic reshaping of the magnitude and direction power flows. 
We will focus instead on the condition under which, in presence of distributed generation,
the system can be brought beyond its design parameters and subject to cascading failures. 

To model distributed renewable sources, we will introduced a probability distribution 
of load demands representing fluctuations due to consumer demand and to an erratic distributed 
generation. Our model is a crude model of reality that ignores effects like the correlations between
different consumers or distributed producers (due for examples to weather conditions) or 
time correlations (like the time-cyclic components of the demand).

To sample the effects of  stochastic demand and production, we will employ intensive Monte Carlo methods 
in which possible power flows configurations are repeatedly calculated during cascading events; 
therefore, we will use the less computational demanding DC power flow model as the building 
block of our simulations. We will then analyze the redistribution of flows due to branch failures 
and the subsequent fragmenting of the power grids caused by overloading cascades.

\section{Methods}

\subsection{Power Flow models}

The AC power flow is described by a system of non-linear equations
that allow to obtain complete voltage angle and magnitude information
for each bus in a power system for specified loads \cite{GraingerBOOK1994}.
A bus of the system is either classified as Load Bus if there are
no generators connected or as a Generator Bus if one or more generators
are connected. It is assumed that the real power $PD$ and the reactive
power $QD$ at each Load Bus are given while for Generator Buses the
real generated power $PG$ and the voltage magnitude $|V|$ are given.
A particular Generator Bus, called the Slack Bus, is assumed as a
reference and its voltage magnitude $|V|$ and voltage phase $\Theta$
are fixed. The branches of the electrical system are described by
the bus admittance matrix $Y$.

The power balance equations can be written for real and reactive power
for each bus; real and reactive power flow on each branch as well as 
generator reactive power output can be analytically determined, 
but due to the non-linear character of the system numerical methods 
are employed to obtain a solution. 

A simplification of the AC power flow equations is obtained by 
linearizing the equations by requiring that bus voltages $V_i$ 
are fixed and phase differences $\theta_{ij}=\Theta_i-\Theta_j$ 
along the branches are small. The resulting linear system of equations 
constitutes the DC power-flow model \cite{GungorBOOK1988}

\begin{equation}
\mathbf{P}=\mathcal{L}\mathbf{\Theta}
\label{eq:DCpf}
\end{equation}

where $P_{i}$ is the total power (generation minus load) at the $i$-th bus, 
$\mathcal{L}=K-Y$ is the Laplacian matrix with $K$ diagonal degree matrix 
$K_{ii}=\sum_j Y_{ij}$; the sign of $P_i$ determines if a node is a generator,
a load or even a transit node ($P_i=0$). Notice that we are neglecting phase
shifts of the transformers that would add an extra term 
in eq.(\ref{eq:DCpf}): $\mathbf{P}=\mathcal{L}\mathbf{\Theta}+\mathbf{P}^\phi$.

\subsection{Overload Cascade model}

To consider the effects of customer behavior and of the 
distributed generation due to erratic renewable sources
like sun and wind in the simplest approximation, 
we  model the effects of ``green generators'' and of variable customer demand 
as a stochastic variation of the power requested by load buses. 

We will assume the case of the full penetration of renewable sources
in the grid; therefore, all the loads will be considered random variables.

A requirement for the stability of the load and generation
is the condition that all branches and buses operate within their
physical feasibility parameters; going beyond such parameters can
trigger cascades of failures eventually leading to black outs \cite{Pahwa2010}.

The redistribution of power is dependent on the electrical
characteristics, such as impedances, of transmission lines.
We neglect line resistances because they are very small 
as compared to their inductive reactances \cite{GungorBOOK1988}.

Flow in power grids has a complex dynamics even in the DC approximation:
if a single branch gets overloaded or breaks, its power is
immediately distributed not to a single different branch but in the whole system.
If the redistribution of power leads to the subsequent overloading of other branches, 
the consequence could be a cascade of overloading failures.
In fact, after an initial failure, some of the branches can get overloaded and
fail: this represents the first stage of cascade. First
stage could possibly lead to overloading and collapse of further branches, 
constituting the second stage and so on. In this way, the system goes 
through multiple stages of cascade until it finally stabilizes and there are 
no more failures. We indicate the final stable configuration by $\lbrace y \rbrace$, where
\begin{equation}
y_{ij}=\left\{ 
	\begin{array}{cc}
	0 & if \, branch\,(i,j)\,is\,broken \\
	Y_{ij} &  otherwise
	\end{array}
\right.
\end{equation}
and $\vert y_{ij} \theta_{ij}\vert<F_{ij}$, where $F_{ij}$ is the threshold
flow beyond which a branch becomes overloaded and fails.

The Overload Cascade Model (OCM) therefore corresponds to a double minimization of 
the objective function
\begin{equation}
H\left(\lbrace y \rbrace\right)=\sum_{ij} y_{ij}^2 \left[ \theta_{ij}^2 - T_{ij}^2\right]
\end{equation}
(here $T_{ij}=F_{ij}/Y_{ij}$ and $y_{ij}\in\lbrace0,Y_{ij}\rbrace$), both respect 
to global variables $\theta_{ij}$ at fixed $y_{ij}$ (via DC power flow) 
and respect to the local variables $y_{ij}$ at fixed $\theta_{ij}$ (breakdown of 
overloaded links). Therefore, the OCM can be mapped in the model for the breakdown of a disordered media 
\cite{ZapperiPRL1997}, indicating that the cascading transition in power grids is a first order 
transition, i.e. consists in an abrupt failure of the system.

\section{Results}

\subsection{AC-DC power flow comparison}

\begin{figure}
\centering
\includegraphics[height=4.cm]{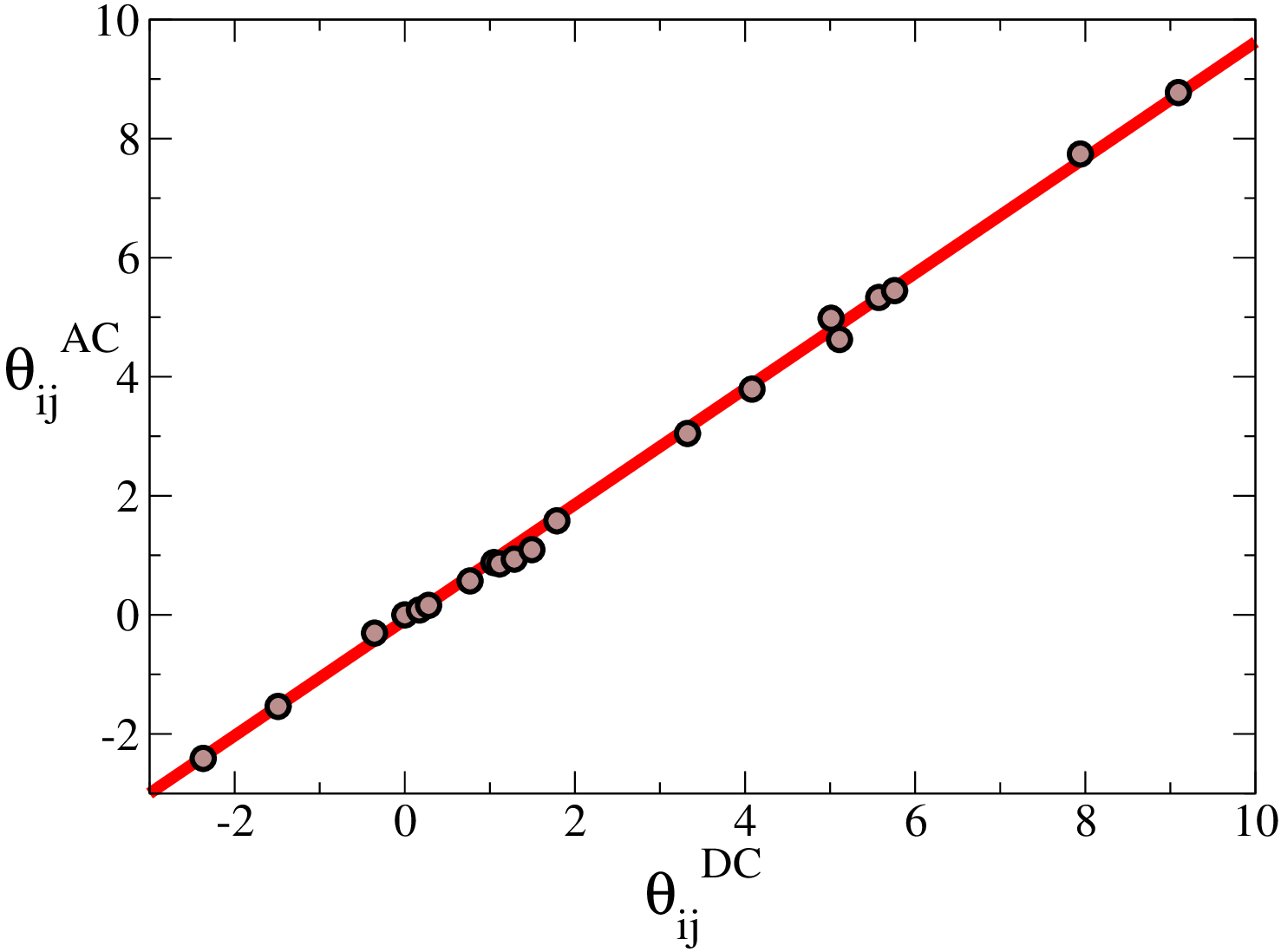} $\,\,\,$ \includegraphics[height=4.cm]{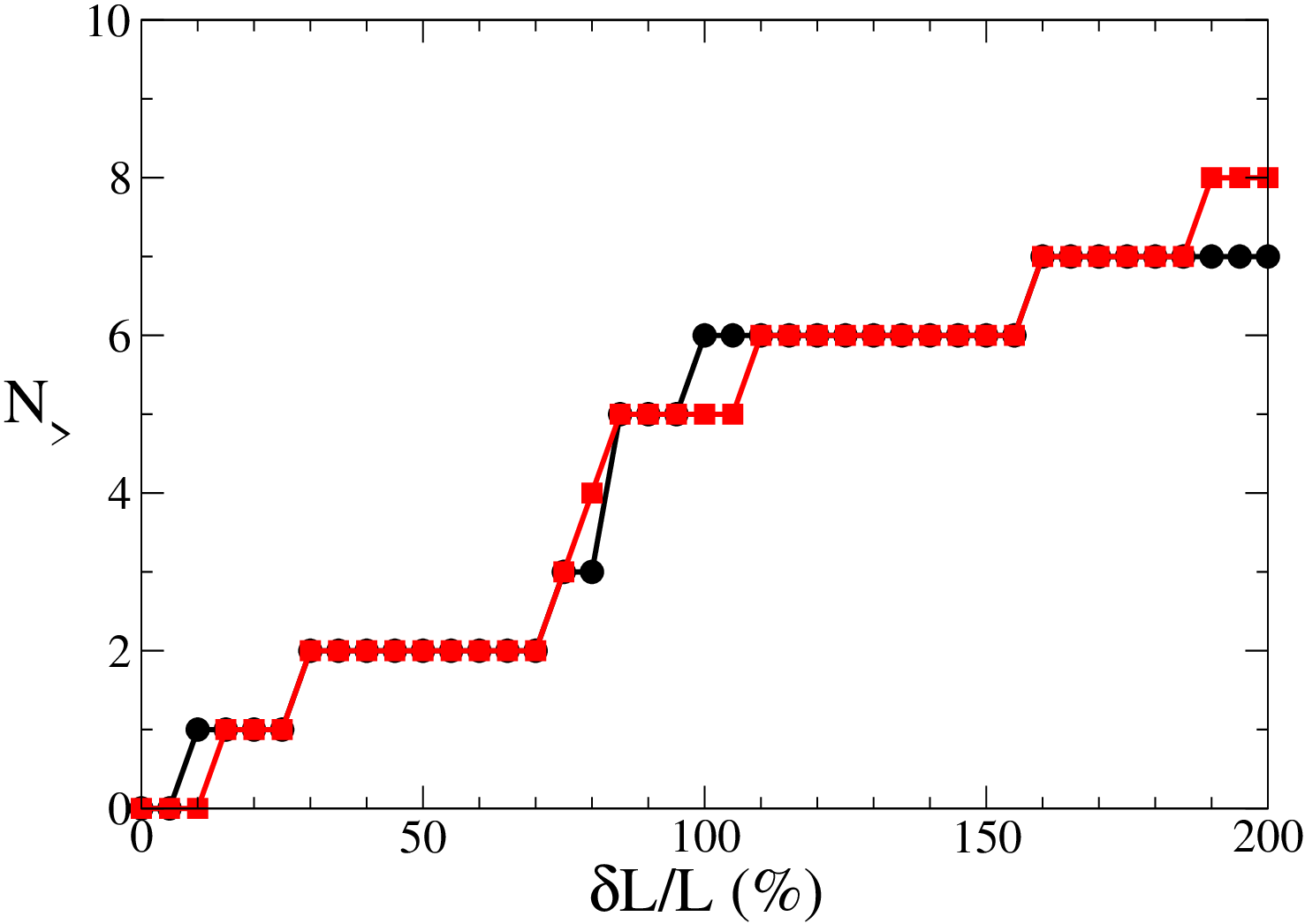}\\
$ $\\$ $\\
\includegraphics[height=4.cm]{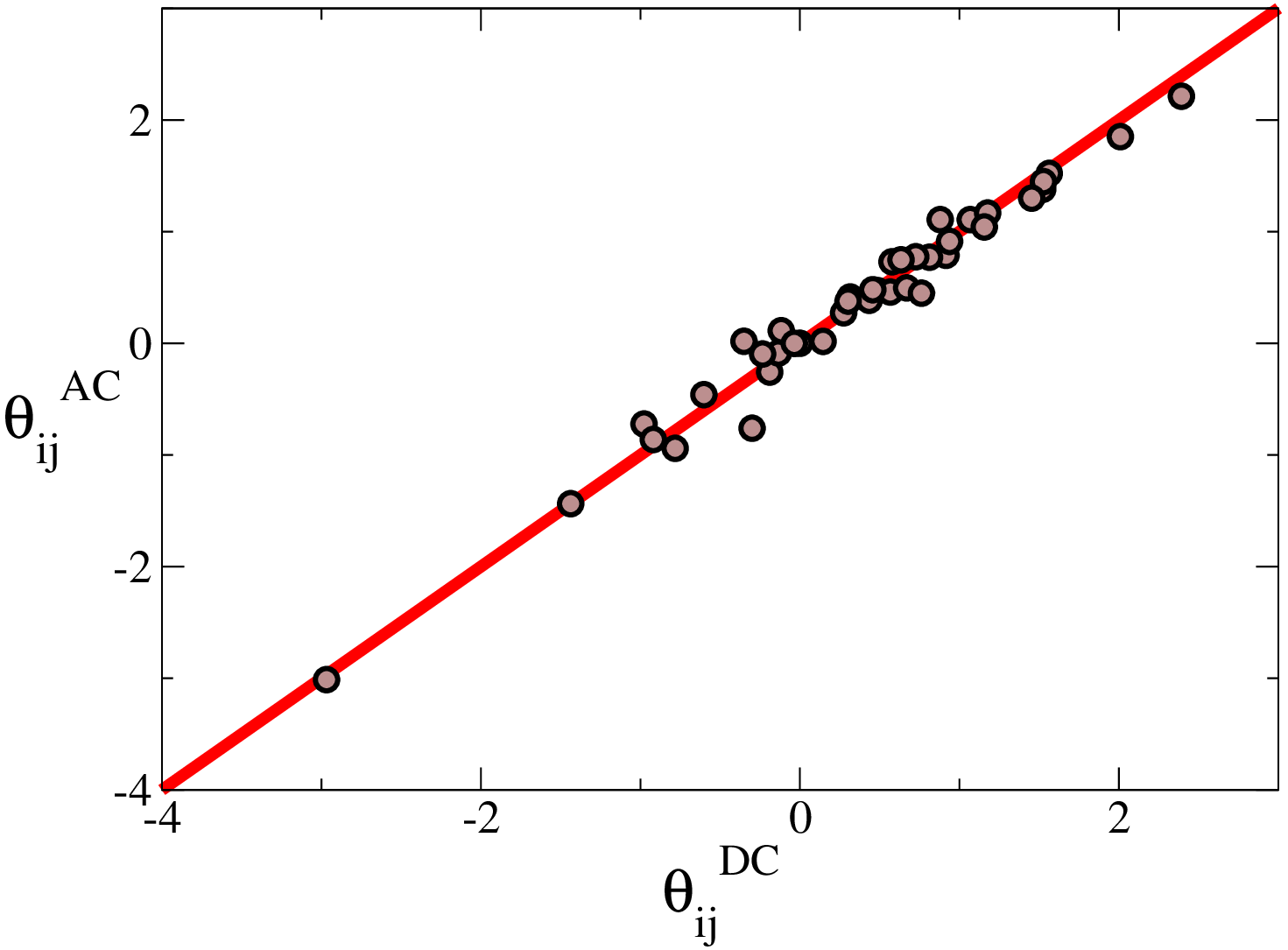} $\,\,\,$ \includegraphics[height=4cm]{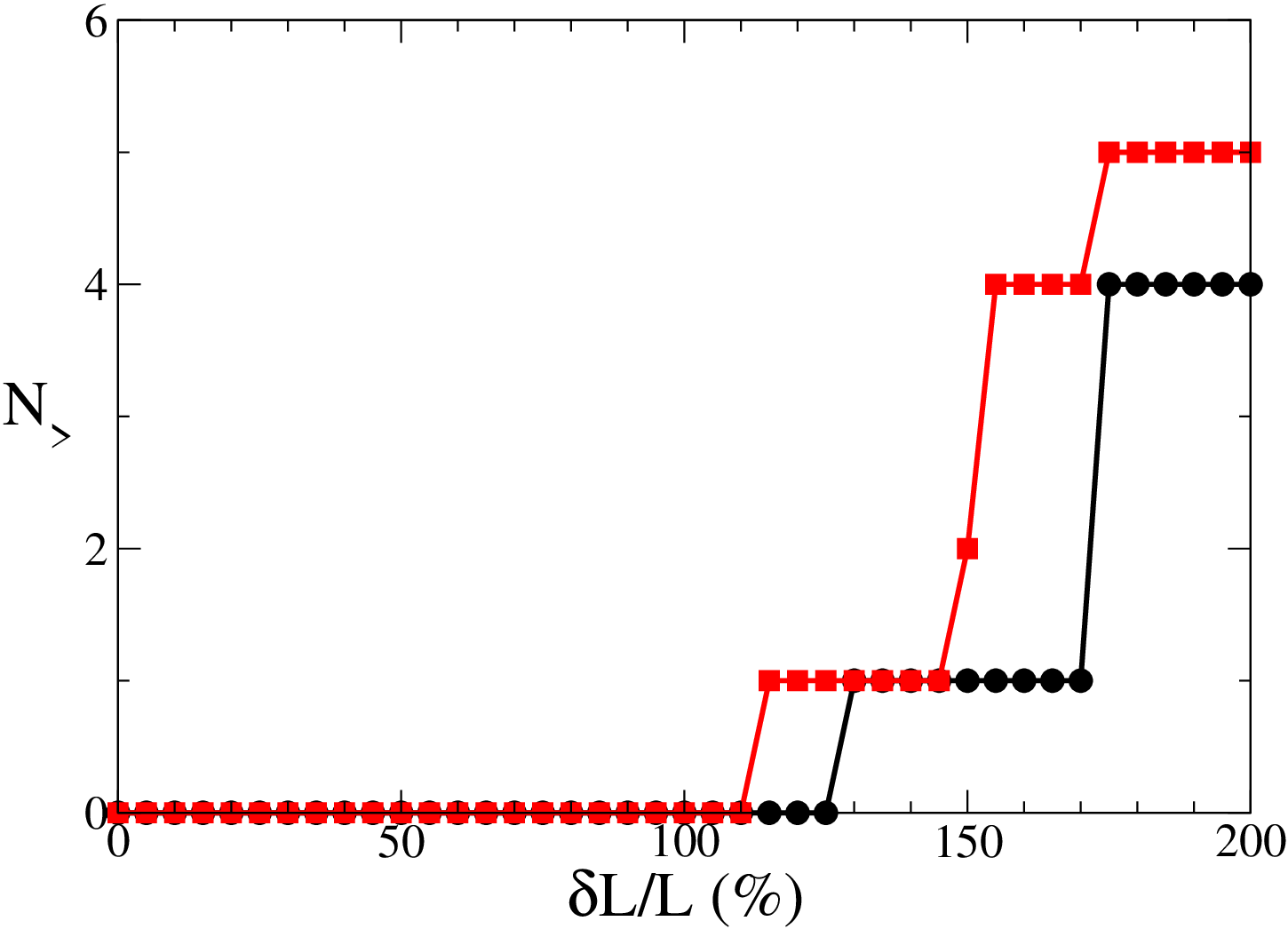}\\
$ $\\$ $\\
\includegraphics[height=4.cm]{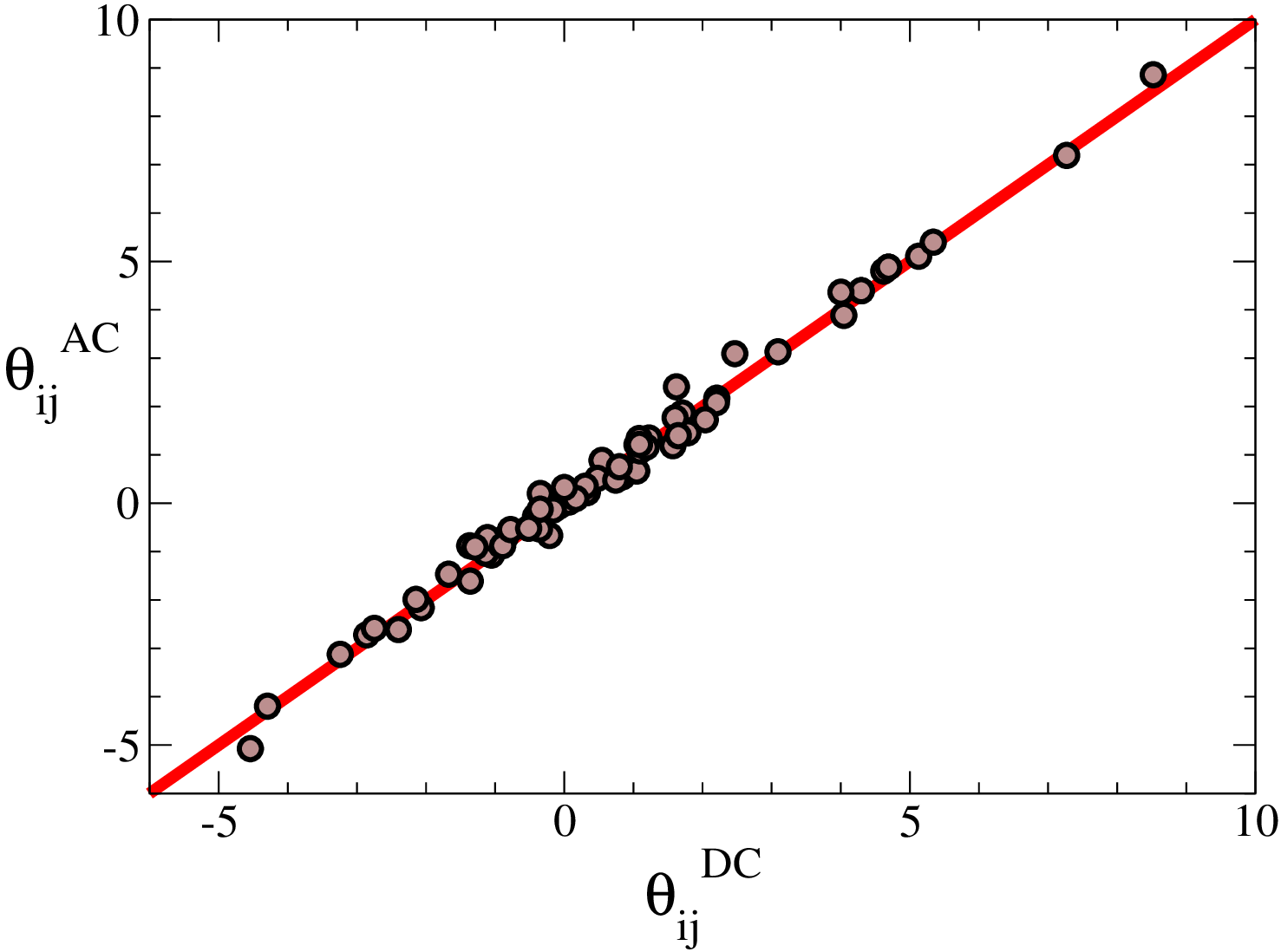} $\,\,\,$ \includegraphics[height=4.cm]{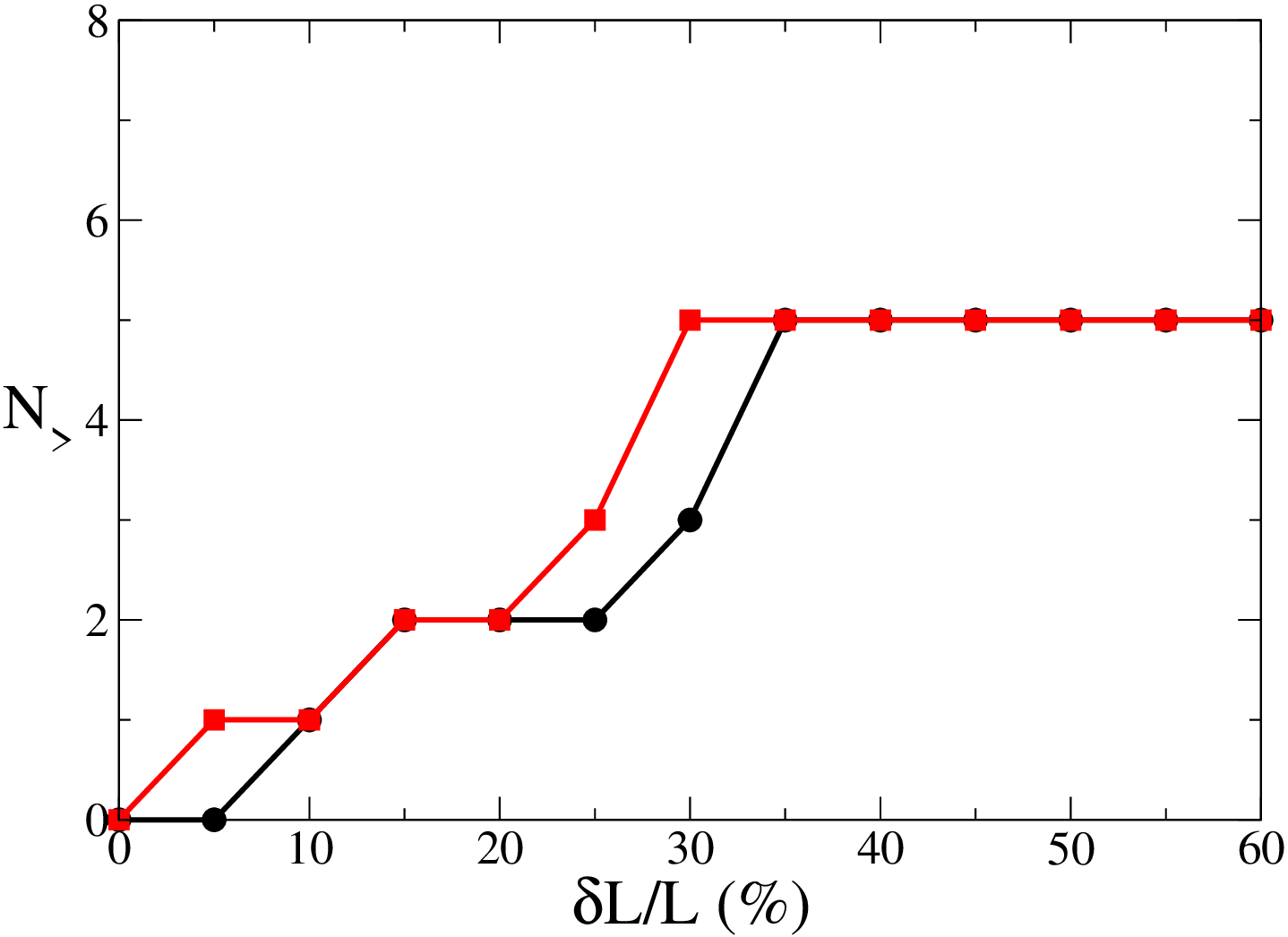}
\caption{
Comparison of DC and AC phase shifts in the $14$, the $30$ and the $57$ bus model grids (top to bottom).
Left panels: The phase angle differences for $AC$ and $DC$ power flows are highly correlated 
$\theta^{AC}_{ij} \sim \theta^{DC}_{ij}$, as seen with the comparison with 
the line $y=x$ (thick straight line).
Right panels:Variation of the number $N_>$ of branches with phase shift above the threshold $\vert\theta_{ij}\vert>10^\circ$ (high flow of power) while varying uniformly the loads by a factor 
$\delta L / L$ respect to the initial condition (where $L$ is the initial total load). Again,
the behavior is equivalent for the DC and the AC models. Notice that the biggest grid ($57$ $buses$) 
becomes unstable even for smaller loads $\delta L / L <0$ (not shown in figure), 
indicating that it is designed to work around a given flow configuration.
} 
\label{fig:DCvsAC}
\end{figure}

The DC model is often used in applications where fast convergence is needed, like in the 
case of real-time systems or when trying to optimize a system calculations must be repeated 
under different conditions. DC power flow is on average wrong by a few percent \cite{Stott2009} 
respect to the more  computational intensive AC power flow. In our case with stochastic sources, 
we want to explore the probability that links get overloaded and fail; therefore, we need a correct 
statistics of the power dissipated along the branches. 
To solve both DC and AC power flow equations, we employ the MATPOWER libraries \cite{matpower} under the 
scientific environment for numerical calculus Octave \cite{octave,octaveBOOK}.

We first compare DC and AC phase shifts $\mathbf{\theta}$ along the branches of 
several grids; the left panels of fig.(\ref{fig:DCvsAC}) shows that the two models produce 
highly similar results for the flows. 

We then compare the number of branches with a high power flux ($\vert\theta_{ij}\vert>10^\circ$) when incrementing/decrementing the loads respect to the initial conditions; the right panel of 
fig.(\ref{fig:DCvsAC}) shows that both the DC and the AC model consistently produce an equivalent number 
of branches above the threshold. 

Notice that the number of high power flux branches increases also when loads are decreased; this is 
an indication of the fact that power grids are designed to work under certain flow configurations and
therefore distributed generation can introduce instabilities and line overloads.

\subsection{Cascades from Correlated Overload}

\begin{figure}
\centering
\includegraphics[height=12.cm]{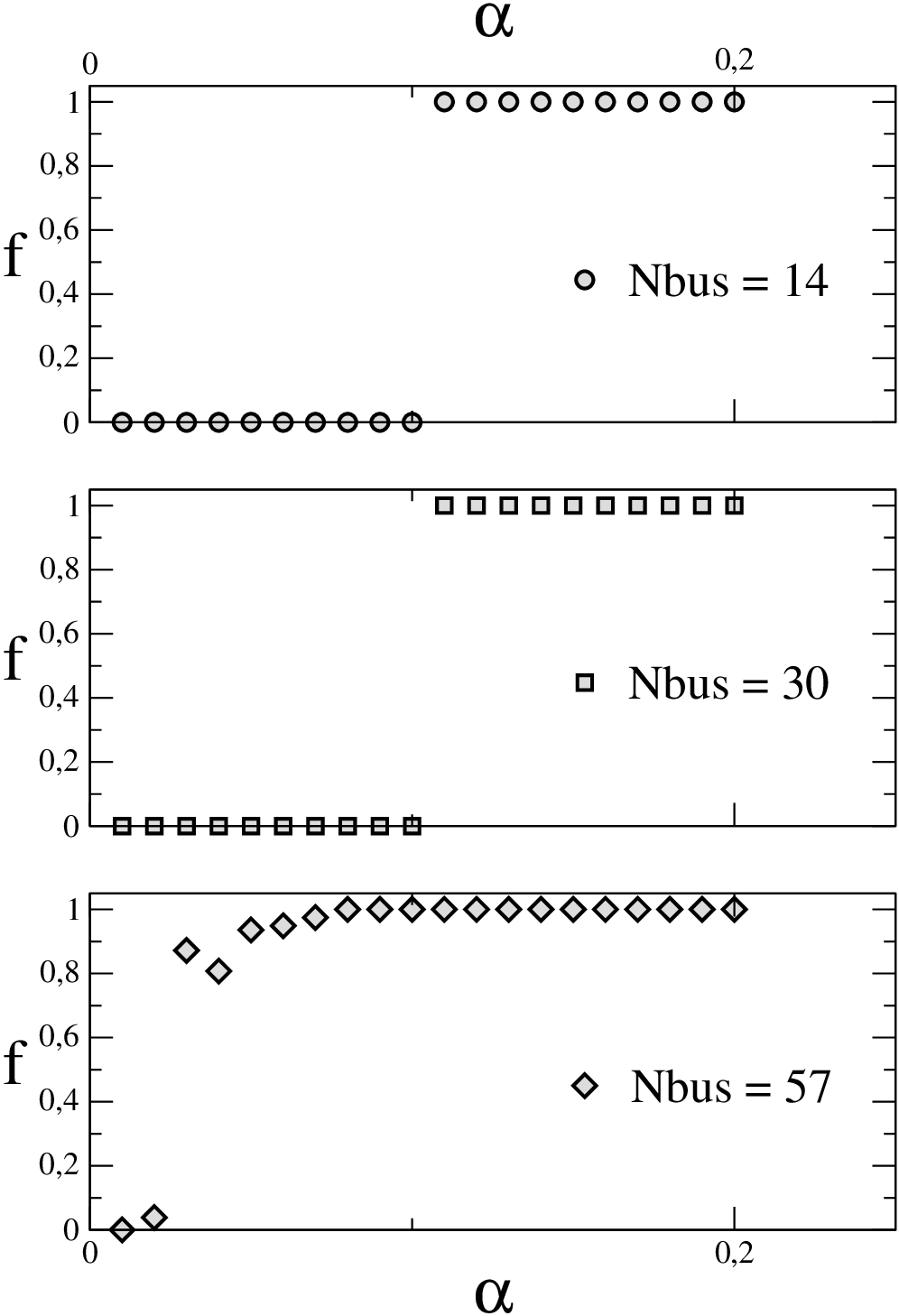}
\caption{
Comparison of the cascade damage from correlated overload for the $14$, the $30$ and the $57$ 
bus model grids (top to bottom). Loads are simultaneously increased by a factor $1+\alpha$
(i.e. $\alpha=0.2$ corresponds to a $20\%$ increase of the loads); notice
that the fraction $f$ of damaged links suffers an abrupt jump as predicted from the mean-field theory.
} 
\label{fig:CorrLoad}
\end{figure}

Customer demand is on average increasing with time, so every kind of human-built network will eventually reach 
its capacity limit beyond which it becomes dis-functional or even breaks down. To mimic such events, 
we analyze via the OCM the effects of a correlated load increase on several model power grids. The loads $L_i$ 
on the $i$-th bus are increased uniformly by a factor $1+\alpha$; in such a situation the mean-field
model of \cite{ZapperiPRL1997} predicts a first order transition (an abrupt jump) in the system. 
Figure (\ref{fig:CorrLoad}) shows that this is indeed the case; notice that up to the transition 
the network stays virtually intact, signaling a very robust design against load increases. On the 
other hand, the fact that in power grids interactions are non-local (due to the non-local character 
of power equations) will eventually lead to a systemic failure for high enough stresses even 
for very well conceived power grids.

\subsection{Effects of distributed generation}

\begin{figure}
\centering
\includegraphics[height=12.cm]{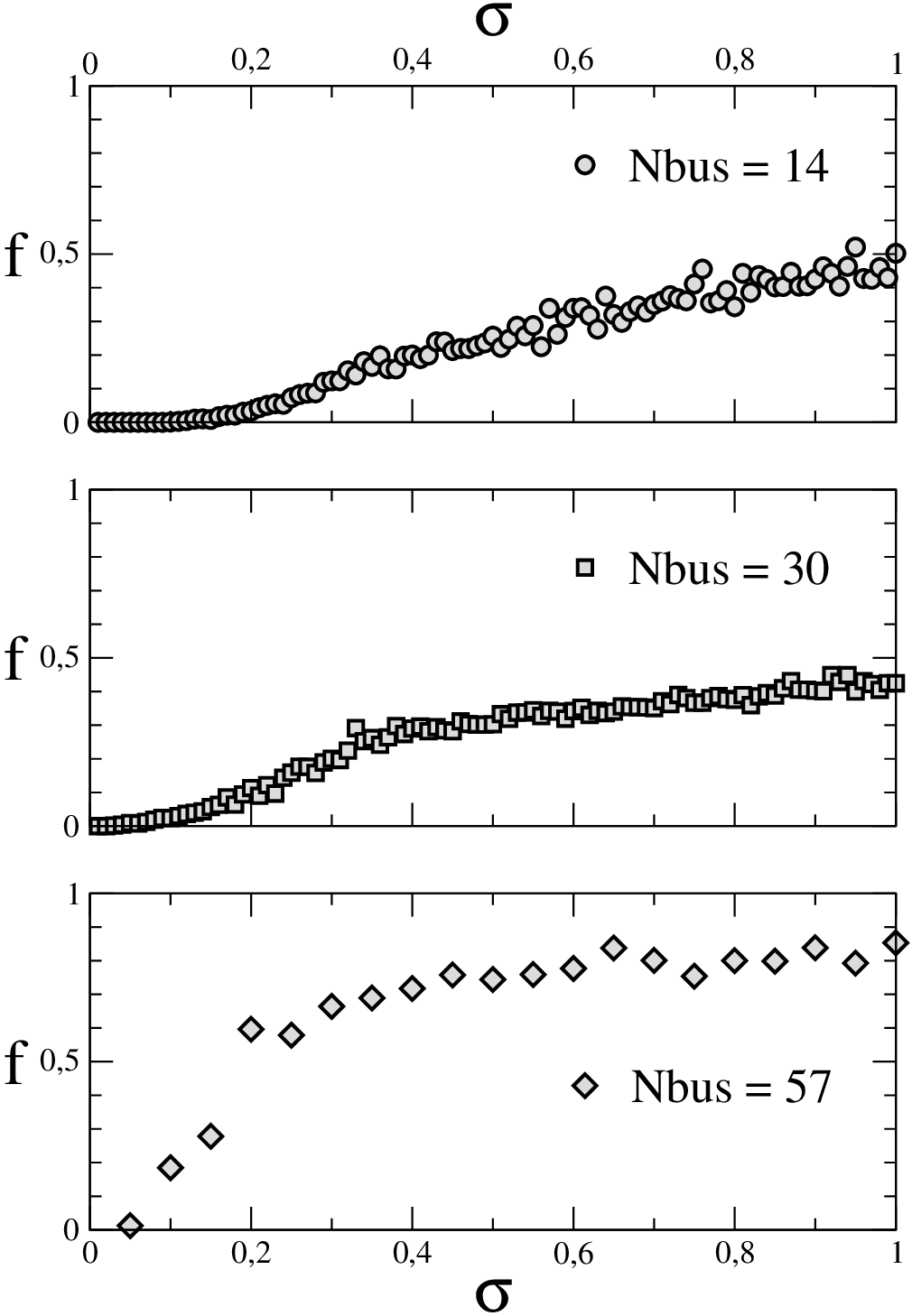}
\caption{
Comparison of the cascade damage from random overloads for the $14$, the $30$ and the $57$ 
bus model grids (top to bottom). The load $L_i$ on each node $L_i$ is subject to a uniform fluctuation 
between $L_i \pm \sigma L_i$ modeling erratic behavior from distributed renewable power sources and customer 
behavior. The fraction $f$ of damaged links has a smoother behavior when compared to the correlated load 
increase case; nevertheless, by increasing the number of nodes in the grid, 
the transition to the failed states becomes steeper, consistently with the first-order 
transition picture of the Overload Cascade Model.
} 
\label{fig:RandomLoad}
\end{figure}

Distributed generation from renewable sources and customer behavior are both sources of randomness
in the power demand/production of a grid. It is therefore important to understand the robustness 
of grids not only respect to variations of the total demand/production but also due to their 
local (i.e. on single buses) fluctuations. To model such issues in the simplest way, we
consider the load $L_i$ on the $i$-th bus to be a random variable uniformly distributed 
in the interval $[L^0_i(1-\sigma),L^0_i(1+\sigma)]$ where $L^0_i$ is the initial load 
and $\sigma$ measures the strength of the fluctuations; 
therefore for $\sigma=1$ each load bus can fluctuate by $100\%$ of its initial value

It is well known that randomness can eventually smooth out transitions \cite{Harris1974}; 
we find that even for the OCM model for random fluctuation transitions are smoother than the 
case of correlated loads. Nevertheless, fig.(\ref{fig:RandomLoad}) also shows that
transitions get steeper when increasing the number of buses as expected from the mean-field character
of the OCM.

\section{Discussion}

We have introduced a model of overload failure cascades based on the DC power flow equation 
that allows to account for the presence of erratic renewable sources distributed 
on a power grid.

While we found that fluctuations increase the instability within an
isolated grid, what happens when more grids are linked together is an open subject.
Power grids are typical complex infrastructural systems; therefore they
can exhibit emergent characteristics when they interact with each other,
modifying the risk of failure in the individual systems \cite{CarrerasHICSS2007}.
As an example, the increase in infrastructural interdependencies
could either mitigate \cite{BrummittPNAS2012}
or increase \cite{LaprieCORR2008,BuldyrevNAT2010} the risk of a system failure.

We find that an increasing uniform stress in power grids can induce abrupt failures
consistently with the universality class of the model of cascade failures we have introduced.
On the other hand, the presence of fluctuations due to erratic renewable sources or to
customer demand induces an apparently smoother break-down of the grid; whether such
transition becomes steeper for bigger grids (as indicated from our simulations
and hinted from the mean-field character of our cascade model) is an important point 
to assess in the near future.

\subsubsection*{Acknowledgements.}
AS acknowledges the support from US grant HDTRA1-11-1-0048, EU FET Open project FOC nr.255987 and
CNR-PNR National Project ”Crisis-Lab”. SP and CS aknowledge the support
of the US Department of Energy grant EE-0003812: “Resourceful Kansas”.
The contents do not necessarily reflect the position or the policy of funding parties.

\bibliography{sesame2012}

\end{document}